\documentclass{article}

\usepackage{PRIMEarxiv}

\usepackage[utf8]{inputenc} %
\usepackage[T1]{fontenc}    %
\usepackage{hyperref}       %
\usepackage{url}            %
\usepackage{booktabs}       %
\usepackage{amsfonts}       %
\usepackage{nicefrac}       %
\usepackage{microtype}      %
\usepackage{lipsum}
\usepackage{fancyhdr}       %
\usepackage{graphicx}       %
\usepackage{svg}
\graphicspath{{media/}}     %

\pdfpageattr{/Group << /S /Transparency /I true /CS /DeviceRGB >>}

\title{Dynaseal: A Backend-Controlled LLM API Key Distribution Scheme with Constrained Invocation Parameters}

\author{
  Jiahao Zhao\textsuperscript{1}, 
  Jiayi Nan\textsuperscript{3}, 
  Lai Wei\textsuperscript{4}, 
  Yichen Yang\textsuperscript{5} \\
  Xi'an University of Posts and Telecommunications \\
  \texttt{\{zjh, nanjy0108, 1606270965, 2647797263\}@stu.xupt.edu.cn} \\
  \And
  Fan Wu\textsuperscript{2} \\
  Xi'an Jiaotong University \\
  \texttt{wfgods@gmail.com} \\
}

\begin{document}
\maketitle

\begin{abstract}

  Due to the exceptional performance of Large Language Models (LLMs) in diverse downstream tasks, 
  there has been an exponential growth in edge-device requests to cloud-based models. 
  However, the current authentication mechanism using static Bearer Tokens in request headers fails to provide the flexibility and backend control required for edge-device deployments. 
  To address these limitations, we propose Dynaseal, 
  a novel methodology that enables fine-grained backend constraints on model invocations.
  \end{abstract}

\section{Introduction}

Large Language Models (LLMs)\cite{hoffmann2022training,kaplan2020scaling}, 
such as ChatGPT\cite{chatgpt}, GPT-4 \cite{GPT-4}, and Claude 3 \cite{Claude_3}, 
have shown remarkable progress and impact across diverse domains\cite{brown2020language}.
Current LLM API access relies on Bearer Token authentication, but this faces challenges with growing edge device inference needs. 
Edge devices include smartphones, PCs, and microcontrollers interfacing with cloud models.

Two common approaches for edge device model invocation are:
\begin{itemize}
    \item \textbf{Pre-embedded API Keys}: API keys are configured in devices, enabling direct model access.
    \item \textbf{Server Relay}: Intermediary servers relay requests, requiring persistent device-server connections.
\end{itemize}

Both approaches have limitations: pre-embedded API keys are vulnerable to security breaches, and server relay introduces latency and bandwidth overhead.

Some attempts have been made by the community to address this issue, but each has its limitations. 
The OpenAI API\cite{OpenAI_Platform_2024} does not provide server-side keys and can only use Bearer Tokens on the client side. 
Zhipu AI's keys\cite{Zhipu_2024} offer both server-side and client-side invocation methods, supporting server-issued keys and expiration control, 
but they cannot restrict critical parameters, leaving them vulnerable to attacks. 
Although OneAPI\cite{songquanpeng_2024} can redistribute keys, the invocation method remains Bearer Token-based, 
failing to resolve the client-side invocation problem.

We propose Dynaseal, a novel methodology that enforces backend model invocation constraints. Users retain token-based authentication simplicity while the backend enforces strict controls on model selection and token limits, enhancing model security.

\section{Related Work}
\subsection{Bearer Token Authentication}
Bearer token authentication is a widely adopted protocol for securing web APIs and services \cite{rfc6750}. This mechanism allows clients to access protected resources by presenting a token, which serves as proof of authorization. The token is typically transmitted in the HTTP Authorization header with the "Bearer" scheme.

\subsection{JWT Token}
JSON Web Token (JWT) represents a compact, URL-safe means of representing claims between parties \cite{rfc7519}. A JWT consists of three parts: a header specifying the signing algorithm\cite{rfc7515}, a payload containing claims, and a signature for verification. The self-contained nature of JWTs eliminates the need for database lookups, making them particularly efficient for stateless authentication. However, this approach also presents challenges in token revocation and session management, requiring additional mechanisms such as blacklisting or short expiration times.

\section{Method}
The system architecture comprises three primary components: Large Language Model (LLM) service providers, backend servers, and edge devices.

\begin{itemize}
    \item The LLM service providers are organizations that host and operate large language models, being responsible for all model-generated responses and inference operations.
    
    \item The backend servers function as authentication endpoints for edge devices while also handling business logic implementations. The specific implementation details are determined by the engineering requirements of each deployment scenario.
    
    \item Edge devices encompass a diverse range of hardware platforms, from sophisticated devices such as smartphones and personal computers to resource-constrained systems like microcontrollers.
\end{itemize}

\subsection{Backend Authentication}
Prior to backend server deployment, a kv-pair (comprising user-id and secret-key) must be obtained from the LLM provider. The kv-pair are subsequently integrated into the service configuration for token generation and identity verification purposes.

\subsection{Token Structure}
\begin{figure}[htbp]
   \centering
   \includegraphics[width=0.8\textwidth]{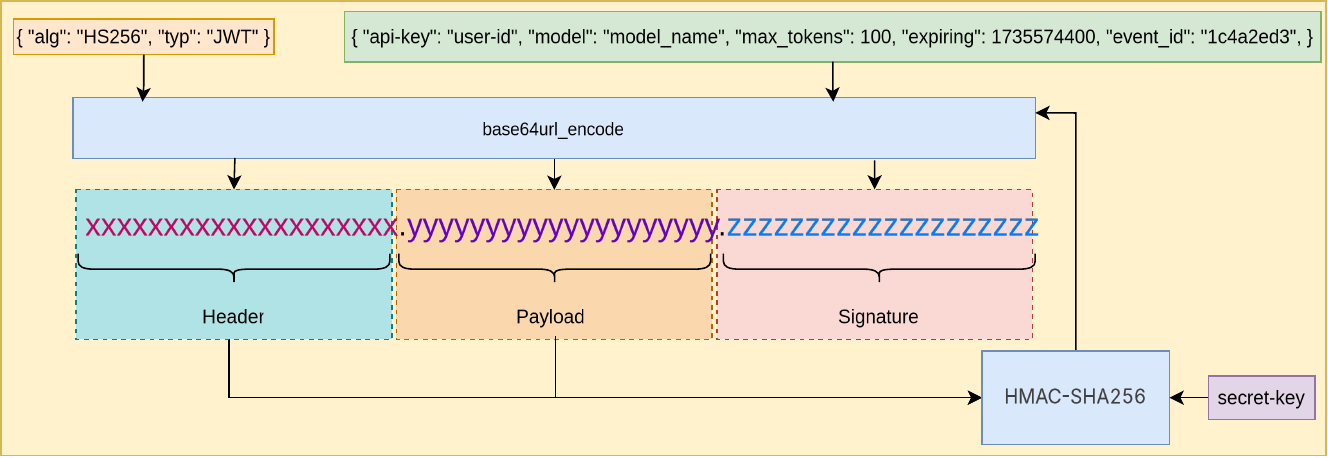}
   \caption{Dynaseal token Structure}
   \label{fig:Dynaseal-token}
\end{figure}

The core mechanism of Dynaseal centers on a specially designed JWT (JSON Web Token) \cite{rfc7519}, with its structural composition illustrated in Figure \ref{fig:Dynaseal-token}.

\begin{itemize}
   \item \textbf{Header}: Declare the encryption algorithm and token type.
   \item \textbf{Payload}: Include key parameters such as model name and maximum token count. The api-key field in payload is configured as user-id of the key-value pair to identify the backend server's identity with the large model provider. The expiration time is set extremely short (e.g., 1s) to prevent reuse.
   \item \textbf{Signature}: Sign with a key-value pair secret-key to ensure token integrity, preventing token tampering.

\end{itemize}

\subsection{Interaction Process}

\begin{figure}[htbp]
   \centering
   \includegraphics[width=0.8\textwidth]{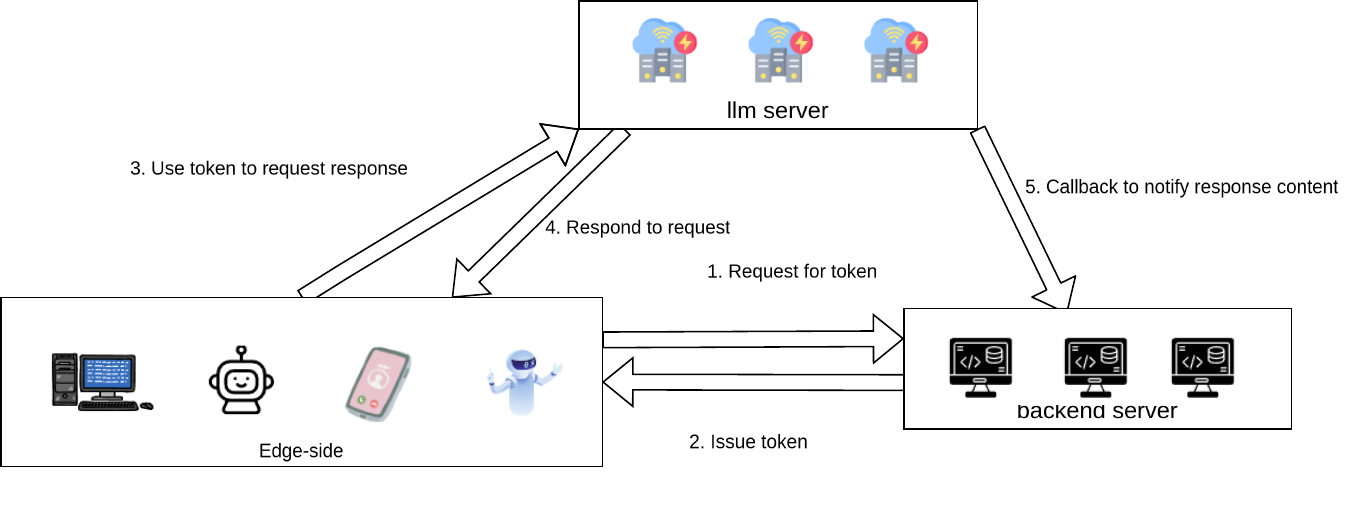}
   \caption{Dynaseal token}
   \label{fig:Interaction-Flow-Diagram}
\end{figure}

As shown in the \ref{fig:Interaction-Flow-Diagram}, the backend server issues tokens to edge devices, with each token encapsulating critical model invocation parameters. Edge devices leverage these tokens to initiate model calls, while the LLM service infrastructure enforces strict invocation constraints based on the parameters embedded within the tokens. Upon completion of the model response, the backend system receives relevant notifications through established callback mechanisms, facilitating comprehensive request lifecycle management.

\begin{enumerate}
   \item \textbf{Request for token}: Edge-side devices requests token from backend for subsequent interactions accorfind to business logic.
   \item \textbf{Issue token}: Backend issues token to edge-side.
   \item \textbf{Request response}: Edge-side uses token to request response insead of Bearer Token.
   \item \textbf{Respond to request}: Large model provider responds to edge-side.
   \item \textbf{Callback to notify response content}: Upon response completion, callback notifies response content.

\end{enumerate}

\subsection{Attack Prevention}

Our system implements comprehensive security measures to prevent potential attacks:

\begin{itemize}
    \item \textbf{Token Tampering}: Malicious actors may attempt to modify token contents to gain unauthorized access. We prevent this through robust \textbf{digital signatures} that ensure token integrity, making any unauthorized modifications detectable.
    
    \item \textbf{Token Replay}: Attackers might try to reuse previously issued tokens. Our system mitigates this risk by implementing \textbf{extremely short validity periods}, rendering captured tokens unusable after expiration.
    
    \item \textbf{Invalid Model Invocation}: To prevent unauthorized model access or parameter manipulation, tokens contain \textbf{critical execution parameters}. The LLM service provider enforces strict invocation constraints based on these embedded parameters, ensuring all calls comply with specified limitations.
\end{itemize}

\section{Experiments}
\subsection{Security Analysis}
As shown in Table \ref{tab:comparison}, we compared the advantages and disadvantages of Dynaseal and three other model invocation methods. We evaluated aspects such as client-side key control, tamper resistance, critical parameter control, and multi-model support, all of which are supported by Dynaseal.

\begin{table}[!h]
   \caption{Comparison of Different Model Invocation Methods}
   \label{tab:comparison}
   \centering
   \begin{tabular}{lcccc}  %
   \hline
   \textbf{API Provider} & \textbf{Client-side} & \textbf{Anti-} & \textbf{Critical} & \textbf{Multi-model} \\
                   & \textbf{key control} & \textbf{tampering} & \textbf{parameter control} & \textbf{support} \\
   \hline
   Openai API      & No  & No  & No  & No  \\
   Zhipu API       & Yes & Yes & No  & No  \\
   OneAPI          & No  & No  & No  & Yes \\
   \bfseries Dynaseal(Ours)  & Yes & Yes & Yes & Yes \\
   \hline
   \end{tabular}
\end{table}

\subsection{Traffic Consumption Comparison}
We compared the network traffic consumption between LLM service providers and backend servers across different approaches including pre-embedded API keys and server relay, as shown in Table \ref{tab:traffic}. The results demonstrate that our method effectively reduces backend server traffic while maintaining constant LLM service provider traffic, simultaneously ensuring key security.

\begin{table}[htbp]
   \centering
   \caption{Traffic Consumption and Key Deployment Comparison}
   \label{tab:traffic}
   \begin{tabular}{lccc}
   \hline
   Method & \begin{tabular}[c]{@{}c@{}}LLM Provider\\ Traffic\end{tabular} & \begin{tabular}[c]{@{}c@{}}Backend Server\\ Traffic\end{tabular} & \begin{tabular}[c]{@{}c@{}}Client-side Key\\ Pre-deployment\end{tabular} \\
   \hline
   Pre-embedded API Key & Normal & None & Required \\
   Server Relay & Normal & 2x & Not Required \\
   Dynaseal & Normal & Minimal & Not Required \\
   \hline
   \end{tabular}
   \end{table}

\section{Conclusion}

We propose a novel method, Dynaseal, allowing backend constraints on model invocation, effectively addressing existing edge-side model invocation security issues while avoiding server relay waste. We provide a complete design and interaction flow, demonstrating the feasibility of this approach.

\bibliographystyle{unsrt}  
\bibliography{references}

\end{document}